\documentclass[twocolumn,showpacs,preprintnumbers,amsmath,amssymb]{revtex4}
\usepackage{latexsym}
\usepackage{graphicx}
\usepackage{amsmath}

\newtheorem{theorem}{Theorem}
\newtheorem{lemma}{Lemma}

\newcommand{\qed}{\rule{7pt}{7pt}}
\newcommand{\ignore}[1]{}

\def\sM{{\cal M}}

\newcommand{\thmref}[1]{Theorem~$\ref{thm:#1}$}

\newcommand{\lemref}[1]{Lemma~$\ref{lem:#1}$}

\renewcommand{\Pr}[1]{\mathbb{P}\left(#1\right)}
\newcommand{\Ex}[1]{\mathbb{E}\left[#1\right]}

\def\eps{\varepsilon}

\def\be{\begin{equation}}
\def\ee{\end{equation}}
\def\bea{\begin{eqnarray}}
\def\eea{\end{eqnarray}}
\newcommand{\ket}[1]{|\, #1\rangle}

\newcommand{\ketbra}[1]{|\, #1\rangle\langle #1\,|}

\newcommand{\eqnref}[1]{(\ref{eq:#1})}

\usepackage{graphicx}
\usepackage{dcolumn}
\usepackage{bm}

\begin{document}


\title{Efficient Generation of Generic Entanglement}

\author{R. Oliveira}
\email{rob.oliv@gmail.com,oscar.dahlsten@imperial.ac.uk,
m.plenio@imperial.ac.uk} \affiliation{IBM Watson Research Center,
Yorktown Heights, NY, USA 10598.}

\author{O.C.O. Dahlsten and M.B. Plenio}

\affiliation{Institute for Mathematical Sciences, Imperial College London, 53 Exhibition Rd, London SW7 2PG, UK}
\affiliation{QOLS, Blackett Laboratory, Imperial College London,
Prince Consort Rd, London SW7 2BW, UK}

\date{\today}

\begin{abstract}
We find that generic entanglement is physical, in the sense that
it can be generated in polynomial time from two-qubit gates picked
at random. We prove as the main result that such a process
generates the average entanglement of the uniform (unitarily invariant) measure
in at most $O(N^3)$ steps for $N$ qubits. This is despite an
exponentially growing number of such gates being necessary for
generating that measure fully on the state space. Numerics
furthermore show a variation cut-off allowing one to associate a
specific time with the achievement of the uniform measure
entanglement distribution. Various extensions of this work are
discussed. The results are relevant to entanglement theory and to
protocols that assume generic entanglement can be achieved
efficiently.
\end{abstract}


\maketitle
{ \bfseries{\em Introduction}} ---  Entanglement has traditionally been viewed as a fundamental tool
for studies of the foundations of quantum mechanics\cite{genovese}. More recently, the viewpoint of using entanglement
as a resource has also gained prominence; see \cite{Plenio V 05} for a recent review.
 While a great deal of insight into the structure of
two-particle entanglement has been gained, it has become equally
clear that the complexity and diversity of multi-particle
entanglement grows exponentially with the number of particles. It
is thus difficult to imagine a structurally simple theory that
characterizes and quantifies all details of multi-particle
entangled states.  On the other hand one may expect that large
numbers of particles admit a notion of typical entanglement
properties for which a structurally simple theory may be
developed. This intuition gives hope that significant progress can
be made by restricting attention to entanglement properties
that are typical (generic) relative to the  uniform (unitarily invariant) measure, the unbiased distribution
of pure states.
 In this setting it was demonstrated that typically  pure
states of large numbers of spins exhibit maximal bi-partite
\cite{pageandfoong, lubkin, lloyd, hayden} and multi-partite
entanglement \cite{hayden}. This suggests that the exploration of
the entanglement properties of generic states is a promising
approach. \\But a big question mark exists as to whether
statements about generic states relative to the uniform measure are
physically relevant. This is  because the generation of a typical unitary
requires  a sequence of 2-qubit unitaries whose length grows
exponentially in the number of qubits \cite{NC}, even if one
allows for a finite fixed fidelity.  Thus achieving the
uniform distribution to a fixed accuracy requires sequences of
random 2-qubit unitaries that grow exponentially with the size
of the system, and quickly becomes unphysical, see e.g.
\cite{emerson1,emerson2,emerson3,smith}.   One could then argue
that the entanglement properties of generic quantum states are
mathematically sound and interesting but physically
irrelevant, as a system undergoing a randomisation of its state through two-party interactions
would only get close to the uniform measure in an unfeasibly long time.
 On the other hand, entanglement properties
represent a restricted class of physical properties of a quantum
state. Accordingly the faithful reproduction of generic
entanglement properties may be possible with far fewer physical
resources, i.e. 2-qubit gates, than those required for the
 generation of the expectation value for an arbitrary
observable.

It is thus crucial to explore whether generic entanglement
properties can be obtained efficiently, i.e. polynomially in the
number of qubits, using only one- and two-qubit gates. The present
work answers this question positively (c.f. Figure 1).

\begin{figure}[th]
\vspace{-0.3cm} \centerline{
\includegraphics[width=9.5cm, height=6cm]{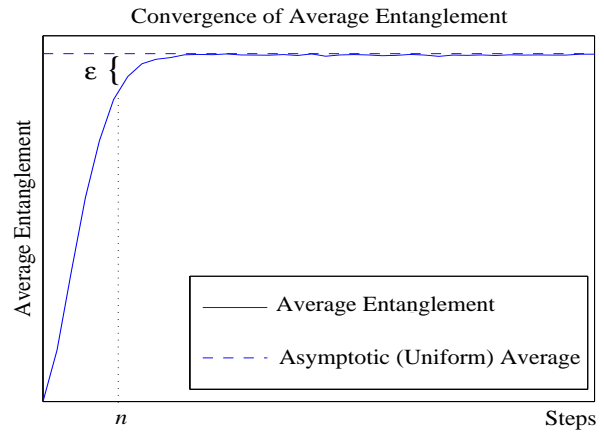}
} \vspace{-0.2cm}\caption{\label{fig:convergence} Typical
numerical simulation using the random circuit. The entanglement
average of the uniform measure is reached to an accuracy $\eps$ in
$n$ steps. We prove that it suffices with $n=O(N^3)$ to achieve a
fixed $arbitrary$ $\eps$ accuracy when increasing $N$.  }
\vspace{-0.3cm}
\end{figure}
\vspace{-0.3cm}

Our results support the physical relevance of the exploration of
generic entanglement towards a structurally  simple entanglement
theory and have direct practical relevance since certain quantum
information processing protocols such as \cite{abeyesinghe,
harrow, hayden, buhrman} assume that generic entanglement can be
generated efficiently.

The presentation proceeds as follows. We firstly define the key
process that is used throughout this work: random two qubit
interactions, modeled as random circuits on a quantum computer.
Then we prove that the generic entanglement average as well as the
purity of a subsystem are achieved efficiently and that the so
generated states are typically very close to maximally
entangled. This is followed by numerical evidence that the
achievement of generic entanglement can be associated with a
specific time, the variation cut-off, for large systems. We finish
with a discussion and conclusion.

{ \bfseries{\em The setting}} -- We consider a set of $N$-qubits split
into two subsets $A$ (with $N_A$ qubits) and $B$ (with $N_B$
qubits). Let $\ket{\psi_0}$ be a initial state in $AB$ and
consider a random circuit $C_n$ consisting of $n$ randomly chosen
two-qubit quantum gates. Define $\ket{\psi_n}=C_n\ket{\psi_0}$ and
the reduced density matrix  $\rho_{A,n}=Tr_B(\ketbra{\psi_n})$ of
system A. Then the entanglement in the state is given by
$E(\psi_n)=S(\rho_{A,n})$ and its purity by $Tr(\rho_{A,n}^2)$.

{ \bfseries{\em Definition of the random circuit}} -- The random circuit $C_n$
is a product $W_n\ldots W_1$ of two-qubit gates where each $W_i$
is independently chosen in the following way: A pair of distinct
integers $c\neq t$ is chosen uniformly at random from
$\{1,\dots,N\}$. Next, single-qubit unitaries $U[c]$ and $V[t]$
acting on qubit $c$ and $t$ respectively are drawn independently
from the uniform measure on $U(2)$. Then $W=CNOT[c,t]U[c]V[t]$
where $CNOT[c,t]$ is the controlled-NOT gate with control $c$ and
target $t$ \cite{gates}.

{ \bfseries{\em Asymptotics of random circuit}} --- The circuit, acting on $N$
qubits, will asymptotically induce the uniform measure on states
(see e.g. \cite{emerson2}). In the above setting for states
distributed according to the uniform measure the average bipartite
entanglement can be found exactly \cite{pageandfoong} and is bounded
from below such that $\Ex{E(\psi_{n})}\geq N_A -\frac{1}{\ln
2}2^{-t}$ where $N_B-N_A=t\geq 0$ \cite{hayden}. Likewise we find
that the average purity of the subsystem $A$ is given by
$(2^{N_A}+2^{N_B})/(2^N+1)$ consistent with \cite{smith}.
Furthermore, the distributions for entanglement and purity
concentrate around their average with increasing $N$
\cite{lubkin,lloyd,pageandfoong,hayden}. Thus one is overwhelmingly
likely to find near-maximal entanglement for large systems.

{ \bfseries{\em Main Theorem}} --- We will now be concerned with the approach
to the asymptotic regime. For the above setting we prove that,
independently of the initial state $|\psi_0\rangle$, convergence
of the expected entanglement to its asymptotic value to an
arbitrary fixed accuracy $\eps$ is achieved after a number of
random two-qubit gates that is polynomial in the number of qubits.
More precisely we find:
\vspace{-0.5cm}
\begin{theorem}\label{thm:main} Suppose that $N_B-N_A=t\geq 0$
and that some arbitrary $\eps\in (0,1)$ is given. Then
for a number $n$ of gates in $C_n$ satisfying
\begin{displaymath}
\vspace{-0.4cm}
    n\geq 9N(N-1)[(3\ln 2)N + \ln\eps^{-1}]/4,
\end{displaymath}
\begin{equation}
    \label{theo1}
\mbox{we have}\;\;\;\;\;\;\;\;\;\;\;\;  \Ex{E(\psi_{n})} \geq N_A - (2^{-t} +\eps)/\ln 2.
\end{equation}
\begin{equation}
    \label{theo2}
  \mbox{and }\;\;\;\;\;\;  \mathbb{E}[\max_{|\Psi\rangle_{AB}= max ent}|\langle\psi_n|\Psi\rangle|]
    \geq 1 - \sqrt{\frac{2^{-t}+\epsilon}{2\ln 2}}.
\end{equation}
\end{theorem}
\vspace{-0.3cm}
 Eq. (\ref{theo2}),  follows from eq.
(\ref{theo1}) employing $\sqrt{2S(\sigma \| \rho)} \geq tr|\sigma-\rho |_1$, where $S$ is the relative entropy, and
$\frac{1}{2}tr|\sigma-\rho |_1 \geq 1 - tr\sqrt{\sqrt{\rho}\sigma \sqrt{\rho}}$
 as well as Uhlmann's Theorem\cite{NC}. To
prove eq. (\ref{theo1}) we prove a Lemma that considers the
quantity $\Ex{Tr(\rho_{A,n}^2)}$.
\vspace{-0.2cm}
\begin{lemma}\label{lem:main} For arbitrary $N,N_A,N_B$ and all $n$ we have
\vspace{-0.2cm}
\begin{displaymath}
\vspace{-0.2cm}
    \left|\Ex{Tr(\rho_{A,n}^2)} - \frac{2^{N_A}+2^{N_B}}{2^N+1}\right|\leq 4^{N}e^{-\frac{4n}{9N(N-1)}}.
\vspace{-0.1cm}
\end{displaymath}
\end{lemma}
 To see that \lemref{main} implies \thmref{main} note first that
$E(\psi_n)=S(\rho_{A,n})\geq -\log_2 Tr(\rho_{A,n}^2)$.
By convexity we then find $\Ex{-\log_2 Tr(\rho_{A,n}^2)}\geq
-\log_2(\Ex{Tr(\rho_{A,n}^2)})$ and a direct computation using
 $\ln (1+x) \le x$ for $x\ge 0$ completes the argument.

\noindent {\em Proof of Lemma 1} --  
We proceed to outline the proof of \lemref{main} below, omitting
some tedious but straightforward calculations to improve clarity. We begin with a useful
representation of quantum states in terms of Pauli-operators.
Indeed, $\ketbra{\psi_n}
=
\sum_{p\in\{0,x,y,z\}^N}\xi_n(p)2^{-N/2}\otimes_{i=1}^N\sigma^{p_i}[i]$,
where each
$\xi_n(p)=2^{-N/2}Tr(\otimes_{i=1}^N\sigma^{p_i}[i]\ketbra{\psi_n})$
 and $\sigma^{p_i}[i]$ is a Pauli operator acting on qubit
$i$ \cite{gates}. Then for the reduced density operator
$\rho_{A,n}=Tr_B(\ketbra{\psi_n})$ we find 
\begin{equation}\label{eq:TraceSq}
   \Ex{Tr(\rho^2_{A,n})} = 2^{N_B}\sum\limits_{\{p\,:\, \forall i\not \in A,\;
    p_i=0\}}\Ex{\xi_n^2(p)}.
\end{equation}
The main purpose will now  be to analyze the evolution of
the expected values of the squared coefficients,
$\Ex{\xi_n^2(p)}$

{ \bfseries{\em Evolution of the coefficients}} --  The key idea of the
proof relies on the observation that the $\Ex{\xi_n^2(p)}$ form a
{\em probability distribution} on $\{0,x,y,z\}^N$ for all $n$ and
that these probabilities evolve as a {\em Markov chain} with
transition matrix $P$ which takes $q$ distributed according to
$(\Ex{\xi_n^2(q)})_q$ in one step to  $p$ distributed
according to $(\Ex{\xi_{n+1}^2(p)})_p$.  To determine $P$ we
consider the action of a random unitary $W_n$ at time $n$ that
acts on qubits $c,t$ in state $|\psi_n\rangle$.  This results
in $\Ex{\xi_{n+1}^2(p)\!\mid \!\psi_n,c,t} \!=\!1/16\!\!\!\!\!\!\!\!\!\sum\limits_{q,q'\in \{0,x,y,z\}^N\!:\!\forall
i\not\in\{c,t\},\,q_i\neq q'_i}\!\!\!\!\!\!\!\!\!\!\!\!\!\!\!\!\xi_{n}(q)\xi_n(q')\!  \times
  \times \Ex{Tr[U\sigma^{q_c}[c]U^\dag \sigma^{\hat{p}_c}[c]]Tr[U\sigma^{q'_c}[c]U^\dag \sigma^{\hat{p}_c}[c]]}\times 
  \times \Ex{Tr[V\sigma^{q_t}[t]V^\dag \sigma^{\hat{p}_t}[t]]Tr[V\sigma^{q'_t}[t]V^\dag \sigma^{\hat{p}_t}[t]]}
$ , where the $(\hat{p}_c,\hat{p}_t)$ are uniquely determined by
$CNOT[c,t]\sigma^{p_c}[c]\sigma^{p_t}[t]CNOT[c,t] \!\!=\!\!
\pm\sigma^{\hat{p_c}}[c]\sigma^{\hat{p_t}}[t]$. Direct calculation
with the uniform measure on $U(2)$ shows that the products of
expectations in the sum vanish unless $q=q'$. Then with the
Kronecker symbol $\delta_{i,j}$ we find $\Ex{\xi_{n+1}^2(p)\mid
\psi_n,c,t}  = \!\!\! \sum\limits_{q\in\{0,x,y,z\}^N}
\!\!\!\!\!\!\!\! P^{(c,t)}(q,p)\!\!\!\prod\limits_{i\not
\in\{c,t\}} \!\! \delta_{q_i,p_i}\xi^2_n(q)$
 where $P^{(c,t)}(q,p)\!\!=\!\!1$ if
$\hat{p}_c\!\!=\!\!\hat{p}_t\!\!=\!\!q_c\!\!=\!\!q_t\!\!=\!\!0$; $P^{(c,t)}(q,p)\!\!=\!\!1/3$ if
$\hat{p}_c\!\!=\!\!q_c\!\!=\!\!0$ and $\hat{p}_t\!,\!q_t\!\!\neq\!\! 0$ or if
$\hat{p}_t\!\!=\!\!q_t\!=\!0$ and $\hat{p}_c\!,\!q_c\!\!\!\neq\!\!\! 0$; and
$P^{(c,t)}(q,p) \!\!=\!\! 1/9$ otherwise. Averaging
$P^{(c,t)}(q,p)\prod_{i\not \in\{c,t\}}\delta_{q_i,p_i}$ over the
$N(N\!-\!1)$ choices of $c,t$ produces the entry $P(q,p)$ of the
transition matrix of the desired Markov chain for
$(\Ex{\xi_n^2(q)})_q$.

{ \bfseries{\em Simplifying the Markov chain}} -- Our aim is the
evaluation of eq. (\ref{eq:TraceSq}) and it turns out that this
can be done via a
simplified Markov Chain. Consider $\{q(n)=(q(n)_1q(n)_2\dots
q(n)_N)\}_{n\geq 1}$ as an $n$-step evolution of our Markov chain
$P$. Then the sets $S(n)=\{i\in\{1,\dots,N\}\,:\, q(n)_i\neq 0\}$,
identifying the nonzero elements of $q(n)$ also form a Markov
Chain. Using $S(n)$ in eq. (\ref{eq:TraceSq}) we find
\begin{equation}\label{eq:Trace2}
    \Ex{Tr(\rho^2_{A,n})} = 2^{N_B}\Pr{S(n)\subset A}.
\end{equation}
Thus we need only to consider the chain $\{S(n)\}_n$.

{ \bfseries{\em Convergence rate of the Markov chain.}} --  As it
turns out, our chain is {\em almost ergodic}: removing the isolated
state $S(n)=\emptyset$, we obtain an ergodic chain on $\Omega =
2^{\{1,\dots,N\}}\backslash\{\emptyset\}$. Since
$\Pr{S(n)=\emptyset}=\Pr{S(0)=\emptyset}=2^{-N}$, determining the
convergence rate to the equilibrium of $S(n)$ on $\Omega$ given by
$\sM(S)=3^{|S|}/(4^N-1)$, $S\in\Omega$ is sufficient for our
purposes. Let $Q=(Q(S,S'))_{S,S'\in\Omega}$ be the transition matrix
of the restricted $S(n)$ chain. It has largest eigenvalue $1$ whose
eigenvector determines the steady state solution $\sM$. The
difference to the second largest eigenvalue, the spectral gap
$\lambda_Q$, bounds the convergence rate to the steady state: for
any initial distribution vector $v$, the component of $Q^n v$
orthogonal to $\sM$ shrinks exponentially fast with $\lambda_Q n$. A
quantitative result is provided in Chapter 2 of \cite{montenegro}
(see Corollary 2.15): since our $Q$ is a reversible chain with
$Q(S,S)\geq 1/2$ for all $S\in\Omega$, we obtain
%
$|\Pr{S(n)\!\!\subset A} \!\!\!- \!\!\!\sum_{\emptyset\neq S\subset
A}\!\!\sM(S)|
\leq\!\!e^{-\lambda_Qn}/\sqrt{\min_{T}\sM(T)}\!\!\leq \!\!
2^N e^{-\lambda_Q n}.$
%
We have $\sum_{\emptyset\neq
S\subset A}\sM(S)\!\!=\!\!(4^{N_A}-1)/(4^N-1)$. Putting back the isolated
state $\emptyset$ into the calculations, applying \eqnref{Trace2}
and noting that $2^{N_B}\leq 2^N$ yields $\left|\Ex{Tr(\rho^2_{A,n})} -
\frac{2^{N_A}+2^{N_B}}{2^N+1}\right|\leq 4^N e^{-\lambda_Q n}.$ All that 
remains is to show that $\lambda_Q\geq 4/9N(N-1)$. We use a
well-known variational principle for $\lambda_Q$\cite{footnote2}:
\begin{equation}
\label{eq:varformula}\lambda_{Q} =
\inf\frac{\sum_{S,S'\in\Omega}\sM(S)Q(S,S')(f(S)-f(S'))^2}{\sum_{T,T'\in\Omega}\sM(T)\sM(T')(f(T)-f(T'))^2},
\end{equation}
where the $\inf$ is taken over non-constant $f:\Omega\to
\mathbb{R}$. This is an application of Raleigh's principle to the
second smallest eigenvalue of $I-Q$, which is precisely $\lambda_Q$.
Eq. (\ref{eq:varformula}) implies that if $R$ is the transition
matrix of a Markov chain on $\Omega$ with same stationary
distribution $\sM$ and $\alpha R(S,S')\leq Q(S,S')$ for all
$S,S'\in\Omega$, then the gap $\lambda_{R}$ of $R$ satisfies
$\lambda_{Q}\geq \alpha\lambda_{R}$. This allows us to estimate
$\lambda_Q$ by {\em comparison} with a simpler chain
\cite{diaconisSC}. Indeed, our $R$ will be transition matrix of
chain $\{B(n)\}$ on $\Omega$ defined as follows: Assume $B(n)=B$ and
choose a $1\leq j\leq N$ uniformly at random. If $j\in B$ and
$|B|\geq 2$, set $B(n+1)=B\backslash \{j\}$ with probability $1/3$
and $B_{n+1}=B$ with probability $2/3$. If $j\in B$ and $|B|=1$, do
nothing. If $j\not\in B$, set $B(n+1)=B\cup\{j\}$.  This is a biased
random walk on the hypercube $2^{\{1,\dots,N\}}$ where transitions
to state $\emptyset$ are suppressed. A coupling argument following
e.g. Chapter 4 of \cite{levinPW} shows that $R$ has a spectral gap
$\ge 1/3N$. Moreover, one can check that $\alpha R(S,S')\leq
Q(S,S')$ with $\alpha = 4/3(N-1)$. It follows that $\lambda_Q\geq
\alpha\lambda_{R}\geq 4/9N(N-1)$, as desired, and the proof is
finished. Numerics indicate convergence in approximately $N\log N$ steps,
so our bound is not tight.

{ \bfseries{\em Observe cut-off}} -- Many Markov chains exhibit the so called
"cut-off effect"\cite{diaconis}. The cut-off refers to an abrupt
approach to the stationary distribution occurring at a certain
number of steps taken in the chain. Say we have a Markov chain
defined by its transition matrix P, and that it converges to a
stationary distribution $\pi$. Initially the total variation
distance $TV=\parallel P-\pi \parallel=sup \mid P(E)-\pi(E) \mid $
between the corresponding probability distributions is given by
$TV=1$. After $k$ steps this distance is given by $TV(k)=
\parallel P^k-\pi \parallel$. A cut-off occurs, basically, if
$TV(k) \simeq 1$ for $k=0,1,2,...a$ and therafter falls quickly
such that after a few steps $TV(k) \simeq 0$. As we increase the
size of the state space, the ratio of the number of steps
during which the abrupt approach takes place and $a$ should vanish
asymptotically. Then we can say that the randomisation occurs at $a$ steps. 
Rigorously, this may be stated as follows
\cite{diaconis}. \emph{ Let $P_n$, $\pi_n$ be Markov Chains on
sets $\chi _n$. Let $a_n$, $b_n$ be functions tending to infinity,
with ${b_n}/{a_n}$ tending to zero. Say the chains satisfy an
$a_n$, $b_n$ cutoff if for some starting states $x_n$ and all
fixed real $\theta$ with $k_n=\lfloor a_n+\theta b_n \rfloor$,
then
$\parallel P_n^{k_n}-\pi_n \parallel\longrightarrow c(\theta)$
with $c(\theta)$ a function tending to zero for $\theta$ tending
to infinity and to 1 for $\theta$ tending to minus infinity.}
Here we observe this behaviour in the entanglement distribution,
a functional of the Markov chain on unitaries given by the random circuit,
and we accordingly term this a cut-off.

\begin{figure}[th]

\centerline{
\includegraphics[width=8.0cm]{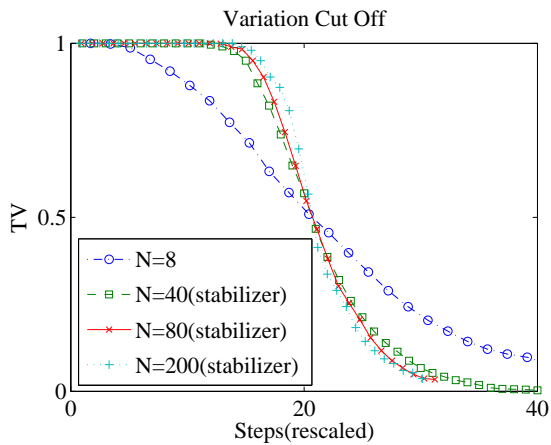}
}\vspace{-0.2cm} \caption{\label{fig:cutoff} Observe a
variation Cut-Off of the entanglement probability distribution
compared with that of the uniform measure as determined numerically.
The state space has been discretised by rounding off entanglement
values to the nearest integer. We observe that $TV\simeq 1$ for a
while and then falls. Finally there is a stage where $TV\simeq 0$.
The effect becomes more pronounced with increasing $N$. The results for $N>8$ are done using the stabilizer random circuit.}
\vspace{-0.4cm}
\end{figure}

\noindent{\bf Numerical Observation}\!\! {\em -- Numerical simulations indicate 
a cut-off effect in the entanglement probability distribution
under the random circuit on $\ket{0}^{\otimes N}$ may be observed
both for single qubit gates drawn from the uniform measure on U(2)
 and for stabilizer gates; see Figure \ref{fig:cutoff}.}

The simulations using stabilizer gates allow us to consider
far larger systems sizes. Here we choose the single qubit
gates $U$ and $V$ from the set
$\{\sigma^x,\sigma^y,\sigma^z,S,H\}$ \cite{gates} with equal
probability. It should be noted that the proof of Lemma 1 still
holds (see \cite{longer} for details)
 and the entanglement behaviour will remain similar \footnote{ C. Dankert, R. Cleve, J. Emerson, E. Livine, arXiv: quant-ph/0606161 which
has since appeared gives further support for this restriction.}. The restriction allows us to use
the efficient stabilizer formalism \cite{gottesman2} and the tools
developed in \cite{audenaert} which in turn allows for an
efficient evaluation of state properties.

{ \bfseries{\em Extensions of the present result --}} Similar methods can be used to address the mixed state setting
through tracing out part of the system on which the random circuit is applied. \ignore{The multipartite entanglement
should also be considered.}  Multipartite entanglement measures based on average purities \cite{Plenio V 05} can be considered with the results established here.
We also anticipate that one can use similar techniques to obtain rigorous statements
about the convergence rates of finite temperature Markov Process Quantum Monte Carlo simulations.

Our results may be applied to the protocols for 
superdense coding of quantum states presented 
in \cite{abeyesinghe, harrow, hayden} to replace 
the inefficient process of creating random unitaries
distributed according to the Haar measure by our 
efficient random circuits. After that replacement, 
Theorem 1 may be applied directly to verify that 
the main Lemma 1 of \cite{harrow} still holds.  It
is an open question whether the performance of the  
protocols in \cite{abeyesinghe, hayden, buhrman} 
is adversely affected by this substitution. This cannot 
be decided on the basis  of Theorem 1 alone but
we expect that similar techniques as described here
and in \cite{longer} will be able to decide this.
The results of this work as well as the above extensions will be
presented in detail in forthcoming publications.

{ \bfseries{\em Conclusion}} --- In this work we have proved that the average
entanglement over the unitarily invariant measure is reached in a
time that is polynomial in the size of the system by a quantum
random process that is restricted to random two-qubit
interactions. We also provided numerical evidence that for large
systems the entanglement distribution of the uniform measure is achieved at a specific point in time,
the variation cut-off. Our results demonstrate that the
entanglement properties of generic entanglement are physical in
the sense that they can be generated efficiently from random
sequences of two-qubit gates. We have described extensions, including how this knowledge can be applied
to render certain protocols efficient.

{ \bfseries{\em Acknowledgements}}  --- 
We gratefully acknowledge early discussions with J.
Oppenheim and discussions with T. Rudolph, G. Smith,
and J. Smolin. M. B. P. was funded by the EPSRC QIPIRC,
The Leverhulme Trust, EU Integrated Project QAP,
and the Royal Society; O. D. by the Institute for Mathematical Sciences 
of Imperial College, and R. O. by NSA and ARDA through ARO
Contract No. W911NF-04-C-0098.


\begin{thebibliography}{foo}
%
\bibitem{genovese} M. Genovese, Phys. Rep. 413,  319 (2005).
%
\bibitem{Plenio V 05} M.B. Plenio and S. Virmani, Quant. Inf. Comp. {\bf 7}, 1 (2006).
%
\bibitem{hayden} P. Hayden, D.W. Leung and A. Winter,
Comm. Math. Phys. {\bf 265}, 95 (2006) .
%
\bibitem{lubkin} E.Lubkin, J. Math. Phys {\bf 19}, 1028 (1978).
%
\bibitem{lloyd} S.Lloyd and H.Pagels, Ann. of Phys. {\bf 188}, 186
(1988).
%
\bibitem{pageandfoong} D.N.Page, Phys. Rev. Lett. {\bf 71}, 1291 (1993).
S.K Foong and S.Kanno, Phys. Rev. Lett. {\bf 72}, 1148 (1994).
%
%
\bibitem{NC} M.A. Nielsen and I. Chuang, {\em Quantum Information and Computation},
Cambridge Univ. Press.
%
\bibitem{emerson2} J. Emerson, E. Livine and S. Lloyd,
Phys. Rev. A. {\bf 72}, 060302 (2005).
%
\bibitem{emerson1} J. Emerson, Y.S. Weinstein, M. Saraceno,
S. Lloyd and D.G. Cory, Science 302, 2098 (2003).
\bibitem{smith} G.Smith and D.W.Leung, Phys. Rev. A 74, 062314 (2006).
%
\bibitem{emerson3}J. Emerson, QCMC04, AIP Conf. Proc. 734, 139 (2004).
%
%
\bibitem{abeyesinghe} A.Abeyesinghe, P.Hayden, G.Smith and A.
Winter, IEEE TIT {\bf 52}, 8, 3635 (2005).
%
\bibitem{buhrman} H. Buhrman, M. Christandl, P. Hayden, H.K. Lo,
and S. Wehner, arXiv:quant-ph/0504078.
%
\bibitem{harrow}
A. Harrow, P. Hayden, and D. Leung, Phys. Rev. Lett., {\bf 92},
187901 (2004).
%
%
\bibitem{emerson4}
J. Emerson, R. Alicki and K. Zyczkowski, J. Opt. B {\bf 7}, 1347 (2005).
%
\bibitem{divincenzo} D.P. DiVincenzo, D.W. Leung, and B.M. Terhal,
IEEE Trans. Inf. Theory, 48(3), 580 (2002).
%
\bibitem{gottesman2} D. Gottesman, arXiv:quant-ph/9807006.
%
\bibitem{audenaert} K.M.R. Audenaert and M.B. Plenio, New J.
Phys. {\bf 7}, 170 (2005). Computer codes downloadable at
www.imperial.ac.uk/quantuminformation.
%
\bibitem{DP} O.C.O.Dahlsten and M.B.Plenio,
Quant. Inf. Comp. {\bf 6}, 527, (2006).
%
\bibitem{diaconis}P.Diaconis, Proc. Natl. Acad. Sci. USA {\bf 93}, 1659, (1996).
%
\bibitem{gates}In the computational basis: $\sigma_x\!\!\!=\!\!\![0,1;1,0],\,\sigma_y\!\!\!=\!\!\![0,-i;i,0],\,\sigma_z\!\!\!=\!\!\![1,0;0,-1],\,S\!\!\!=\!\!\![1,0;0,i],\, 
H\!\!\!=\!\!\![1,1;1,-1]/\sqrt{2}\mbox{ and }CNOT\!\!\!=\!\!\![1,0,0,0;0,1,0,0;0,0,0,1;0,0,1,0]$.
%
%
\bibitem{levinPW} {\em Markov Chains and Mixing Times}
D. A. Levin, Y. Peres and E. L. Wilmer, book draft at http://www.oberlin.edu/markov/
%
\bibitem{diaconisSC} P. Diaconis; L. Saloff-Coste
Ann. of Appl. Prob., {\bf 3}, No. 3., p.696, (1993).
%
\bibitem{montenegro} R. Montenegro and P. Tetali, Ser. Found., Trends, Th. Comp. Sci., {\bf 1:3}, NOW Publ., Boston-Delft, (2006).
%
%
\bibitem{footnote2}See Lemma 2.21 in \cite{montenegro}. Our numerator is twice
their Dirichlet form (Defn. 2.1) and our denominator is twice the
variance.

\bibitem{Briegel} J. Calsamiglia, L. Hartmann, W. D\"ur, and H.-J. Briegel,
Phys. Rev. Lett. {\bf 95}, 180502 (2005).
%
\bibitem{longer} O.C.O Dahlsten, R.Oliveira, M.B.Plenio, arXiv:quant-ph/0701125.
%

\end{thebibliography}
\end{document}